\documentclass[final,5p,times,twocolumn,numbers]{elsarticle}

\usepackage[numbers]{natbib}
\usepackage[superscript]{cite}
\usepackage[hyphens]{url}
\usepackage{graphicx}

\journal{...}

\begin{document}

\begin{frontmatter}




\title{The recessionary pressures of generative AI: A threat to wellbeing}

\author{Jo-An Occhipinti\textsuperscript{a,b,c,*}}
\cortext[cor1]{Corresponding author. Email: jo-an.occhipinti@sydney.edu.au}
\affiliation[first]{Brain and Mind Centre, University of Sydney, Camperdown, Australia}
\affiliation{Mental Wealth Initiative, University of Sydney, Camperdown, Australia}
\affiliation{Computer Simulation & Advanced Research Technologies (CSART), Sydney, Australia}

\author{Ante Prodan\textsuperscript{a,b,c,d}}
\affiliation{School of Computer, Data and Mathematical Sciences, Western Sydney University, Sydney, Australia}

\author{William Hynes\textsuperscript{b,e,f}}
\affiliation{The World Bank, Washington D.C., USA}
\affiliation{Santa Fe Institute, Santa Fe, New Mexico, USA}

\author{Roy Green\textsuperscript{g}}
\affiliation{University of Technology Sydney, Broadway, Sydney, Australia}

\author{Sharan Burrow\textsuperscript{h}}
\affiliation{Visiting Professor in Practice, London School of Economics Grantham Institute, London, UK}

\author{Harris A. Eyre\textsuperscript{b,i,j,k}}
\affiliation{Brain Capital Alliance, San Francisco, California, USA}
\affiliation{Baker Institute for Public Policy, Rice University, Houston, Texas, USA}
\affiliation{Meadows Mental Health Policy Institute, Dallas, Texas, USA}

\author{Adam Skinner\textsuperscript{a}}

\author{Goran Ujdur\textsuperscript{a,b,c}}

\author{John Buchanan\textsuperscript{a,b}}
\affiliation{Business School, University of Sydney, Sydney, Australia}

\author{Ian B. Hickie\textsuperscript{a,b}}

\author{Mark Heffernan\textsuperscript{d,m}}
\affiliation{Dynamic Operations, Sydney, Australia}

\author{Christine Song\textsuperscript{a,b}}

\author{Marcel Tanner\textsuperscript{n,o}}
\affiliation{Swiss Academies of Arts and Sciences, Bern, Switzerland }
\affiliation{Swiss Tropical and Public Health Institute & University of Basel, Switzerland}





\begin{abstract}
Generative Artificial Intelligence (AI) stands as a transformative force that presents a paradox; it offers unprecedented opportunities for productivity growth while potentially posing significant threats to economic stability and societal wellbeing. Many consider generative AI as akin to previous technological advancements, using historical precedent to argue that fears of widespread job displacement are unfounded, while others contend that generative AI’s unique capacity to undertake non-routine cognitive tasks sets it apart from other forms of automation capital and presents a threat to the quality and availability of work that underpin stable societies. This paper explores the conditions under which both may be true. We posit the existence of an AI-capital-to-labour ratio threshold beyond which a self-reinforcing cycle of recessionary pressures could be triggered, exacerbating social disparities, reducing social cohesion, heightening tensions, and requiring sustained government intervention to maintain stability. To prevent this, the paper underscores the urgent need for proactive policy responses, making recommendations to reduce these risks through robust regulatory frameworks and a new social contract characterised by progressive social and economic policies.  This approach aims to ensure a sustainable, inclusive, and resilient economic future where human contribution to the economy is integrated with generative AI to enhance the Mental Wealth of nations. 

\end{abstract}


\end{frontmatter}




\label{introduction}
\begin{center}
\section*{“Why then, can one desire too much of a good thing?”}
\end{center}
\begin{flushright}
— Shakespeare, 1623
\end{flushright}
Stable, quality employment is the bedrock of societal strength, providing not just economic stability but also a source of shared purpose, connectedness, and psychological fulfillment. Work is the steady pulse that sustains individual lives, families, the vitality of communities, and the prosperity of nations. Exclusion from stable, high-quality work can precipitate poor psychological wellbeing with its most severe consequence being the potential for deaths of despair.\cite{RN1,RN2} Guarding against economic downturns and the associated risks to employment is therefore a shared concern for economists, governments, business leaders, labour organisations, \textit{and} mental health researchers. Enhancing productivity has long been a key focus, not only for preventing economic downturns, but to improve standard of living, enhance competitiveness, foster innovation, and create a more resilient labour market. However, since the 2008 global financial crisis, there has been a global slowdown in productivity growth affecting about 70\% of advanced and developing economies.\cite{RN3} In this context, the launch of generative Artificial Intelligence (AI), particularly of large language models like GPT-3 by OpenAI in November 2022 opened up the possibility of a new frontier in generalised productivity growth. Alongside this prospect has been debate on whether AI-driven productivity growth represents an opportunity or a threat to social and economic wellbeing. This paper explores the question: in realising the full potential for AI-driven productivity growth, could too much of a good thing be damaging? We argue that generative AI holds the capacity to profoundly reshape labour market dynamics and paradoxically, if left to market dynamics, undermine the very economic growth it aims to achieve.

Many consider generative AI as akin to previous technological advancements such as the internet or computers and emphasize this development as simply a continuity of the historical pattern of technology-driven progress. They argue that, like past innovations, generative AI is another form of capital that can augment human capabilities to enhance labour productivity, efficiency, economic growth, and hence societal wellbeing. Their perspective contends that fears of widespread job displacement are unfounded, using historical precedent and analyses to demonstrate that technological advancements typically resulted in the creation of new types of jobs, accompanied by a gradual adjustment in the tasks performed by workers, the requisite skill levels, and the share of income earned, rather than an aggregate reduction in the amount of work available.\cite{RN4, RN5} Certainly, the post-pandemic reductions in unemployment rates and increases in skills shortages in industrialised nations are giving credence to this perspective.\cite{RN6} An additional fundamental assumption that underpins this optimistic business-as-usual scenario is that the new jobs that are inevitably created as a result of this technological disruption will require human input, and at a level to offset aggregate job loss. The World Economic Forum’s Future of Jobs Report, 2023, estimates that the largest drivers of job growth (to offset job displacement) will likely be in the areas of big data analytics, climate change and environmental management technologies, encryption and cybersecurity, education, agricultural professionals, and e-commerce and trade.\cite{RN7} While this analysis has some merit, it is too limited in scope and reveals a fundamental misunderstanding of the capabilities of generative AI.

Generative AI is not simply automation, it is a form of intelligence rapidly evolving to acquire experiential learning, creativity, adaptability, strategic thinking, and the ability to generate knowledge surpassing human capacities. GPT-4 already has capability to ‘solve novel and difficult tasks that span mathematics, coding, vision, medicine, law, psychology and more, without needing any special prompting.\cite{RN8} On professional and academic exams it exhibits human-level performance scoring in the top 10\% of test takers in a simulated version of the Multistate Bar Examination.\cite{RN8, RN9} More important to productivity and innovation than human performance in any given area, is the ability to perform across multiple disciplines (e.g. engineering, medicine and law) where human intelligence is constrained due to the sheer amount of information it has to manipulate. OpenAI are developing GPTs that are enhancing productivity not only in generalised everyday tasks but also in professional assignments without the necessity of programming skills. Estimates from the Brookings Institution indicate that, in the coming decade, around 60\% of job tasks in the United States alone are at medium to high risk of being replaced by AI. \cite{RN10} Progress is being made at unprecedented speeds. On average, across a broad range of task categories, a 25\% performance improvement has been achieved with GPT-4 (released March 2023) compared to GPT-3.5 released months earlier.\cite{RN11} While timelines remain uncertain, substantial investments are being made (anticipated to surpass US\$300 billion in 2026 \cite{RN12})  to progress capability, reliability, and implementation of generative AI, including the potential for self-advancement with minimal future human input. Those that understand the capability and trajectory of generative AI see the current surge in jobs for those with specialised AI skills as containing the seeds for their own obsolescence.\cite{RN13, RN14} The pioneers of this technology are now openly acknowledging that generative AI is fundamentally a labour replacing tool.\cite{RN15} 

To appreciate the likely impact of generative AI on the nature of work and the economy it is helpful to first consider historic trends in four occupational categories; routine manual, non-routine manual, routine cognitive, non-routine cognitive (see Appendix Table 1 of Borland \& Coelli, 2023).\cite{RN4} Figure 1 illustrates a typical example of these trends based on the Australian context but similar across other advanced nations. The share of employment consisting primarily of routine manual and routine cognitive tasks have declined with increasing automation, particularly with the rise in IT-enabled technologies from the 1990s.\cite{RN4} However, offsetting this decline has been a 12.3 percentage point increase in the share of employment in non-routine cognitive occupations between 1986 to 2022 demonstrating the creation of new higher skilled, higher paid jobs where the skills required are unique to labour and cannot be replaced by capital.\cite{RN4} The departure from this historical trend comes with the advent of a form of capital (generative AI) that already has the capacity to handle non-routine cognitive tasks.  Unlike earlier automation technologies that were primarily suited for routine and repetitive manual and cognitive tasks, generative AI can analyse complex data, generate creative content, and most recently has demonstrated the capacity to perform other non-routine functions at or above human level performance.\cite{RN8} 

\begin{figure}[ht]
\centering
\includegraphics[width=0.5\textwidth]{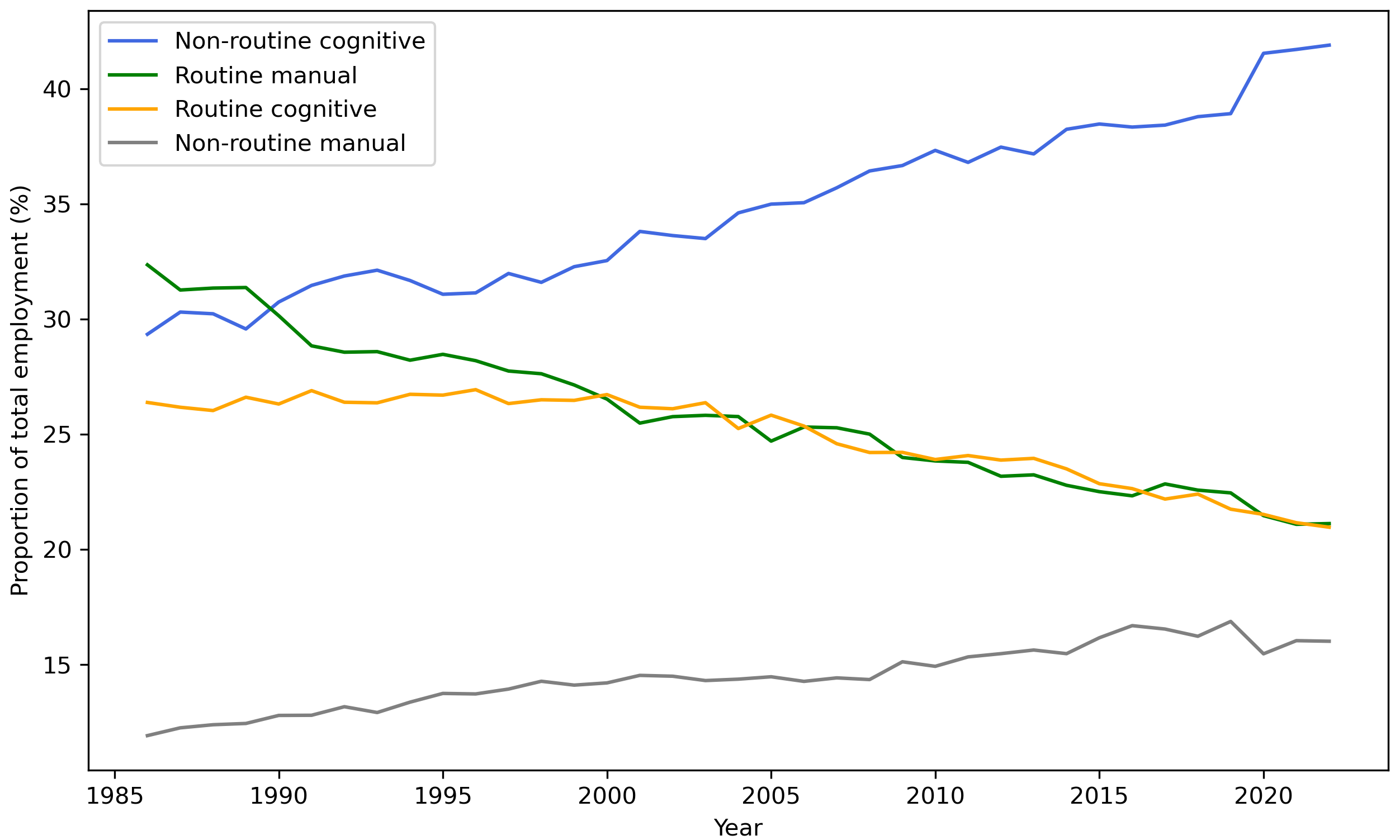}
\caption{Share of employment by type of occupation, Australia, 1986 – 2022 (August) (reprinted with permission. Source: Borland, J. \& Coelli, M. The Australian labour market and IT-enabled technological change. Working Paper No. 01/23. Melbourne Institute of Applied Economic \& Social Research, Melbourne, Australia, 2023).}
\label{fig:image1}
\end{figure}

Non-routine cognitive occupations in this classification include managers, professionals, and technicians.\cite{RN4} A broad range of professional and technical occupations have been identified as facing medium to high exposure to displacement by AI in the coming decade including accountancy, finance, banking, legal, human resources, teaching, data analyst, creative writing, marketing and communications, journalism, and medical diagnostic roles to name a few.\cite{RN16} While it is unlikely that AI will eliminate labour in affected occupations entirely, the need for human input will be dramatically reduced, even for new non-routine cognitive jobs and sectors that may be created as a result of this technological disruption.\cite{RN17}  Occupations characterised by routine cognitive tasks (such as sales and office/administration roles), and some aspects of manual categories are also likely to be increasingly affected by generative AI. While the deployment of generative AI may indeed create new high-skilled occupations where labour is preferred to capital such as those related to AI ethics, data privacy, and algorithm oversight, opportunities in these new occupations will not be sufficient to offset the large-scale displacement of workers across the cognitive occupations. With contraction of the labour market, particularly in the cognitive occupational categories, increases in unemployment, underemployment, non-participation in the labour market, and displacement to lower paid manual jobs (including protective services, food/cleaning, care services, labourers etc) should be expected in the absence of regulation, placing downward pressure on demand for goods and services in the economy.

\label{Recessionary pressures }
\section*{Recessionary pressures }
The amount of downward pressure on demand will depend on the extent of the shift in the ratio of AI-capital to labour (i.e. the extent of job displacement) that is permitted. Replacement of existing capital with AI-capital or smaller increases in the aggregate ratio of AI-capital to labour across the economy are likely to simply augment labour productivity, lowering prices, increasing consumers’ real incomes, increasing demand for goods and services, and creating employment opportunities. However, a threshold likely exists beyond which the substitution of labour for AI-capital (capital deepening) will trigger the following self-reinforcing loop (in the absence of government intervention) depicted visually by a causal loop diagram in Figure 2:
\begin{itemize}
  \renewcommand{\labelitemi}{$\diamond$} 
  \item 	Decreasing availability of higher paid high- and middle-skilled jobs will displace workers from a broad range of professions. This will increase competition for lower-skilled jobs, lowering wage rates (wage compression), reducing job quality, and likely expanding the insecure and exploitive gig economy. 
  \item Average household income will fall, reducing demand for goods and services, thereby exacerbating labour market contraction, and placing further downward pressure on demand. 
  \item Even among workers who retain high-skilled jobs, rising job insecurity will undermine consumer confidence leading to reduced discretionary spending. This shift in consumption patterns would further exacerbate recessionary pressures.
  \item 	A broad range of businesses would face challenges in maintaining profitability, investment would decline, and financial markets are likely to experience greater volatility.
  \item Labour share of income (already in decline since the mid-1980s and attributed largely to the impact of technology)\cite{RN18} will decrease further as owners of AI-capital secure a growing portion of national income.
  \item Increasing income inequality, job scarcity, and associated population displacement, coupled with a contraction in brain capital and diminishing public trust in the economic system, is likely to intensify polarization and political unrest, posing additional challenges to social and economic stability. 
  
\end{itemize}
	
Generative AI enables an aggregate decoupling of economic productivity from labour and its virtualised infrastructure can scale quickly (in days instead of years). While this would theoretically permit rapid and continuous productivity growth, in practice substantial increases in the AI-capital-to-labour ratio needed to achieve that continuous productivity growth would likely fundamentally change labour market dynamics, undermine economic stability, and be detrimental to individual and social wellbeing. Figure 3 provides a stylised graph of the theorized relationship between the AI-capital-to-labour ratio, productivity, and demand, suggesting the potential presence of a threshold beyond which demand falls, and highlighting a ‘productivity overhang’ (the productivity that cannot be realized due to bounded demand for goods and services). 

\begin{figure}[ht]
\centering
\includegraphics[width=0.5\textwidth]{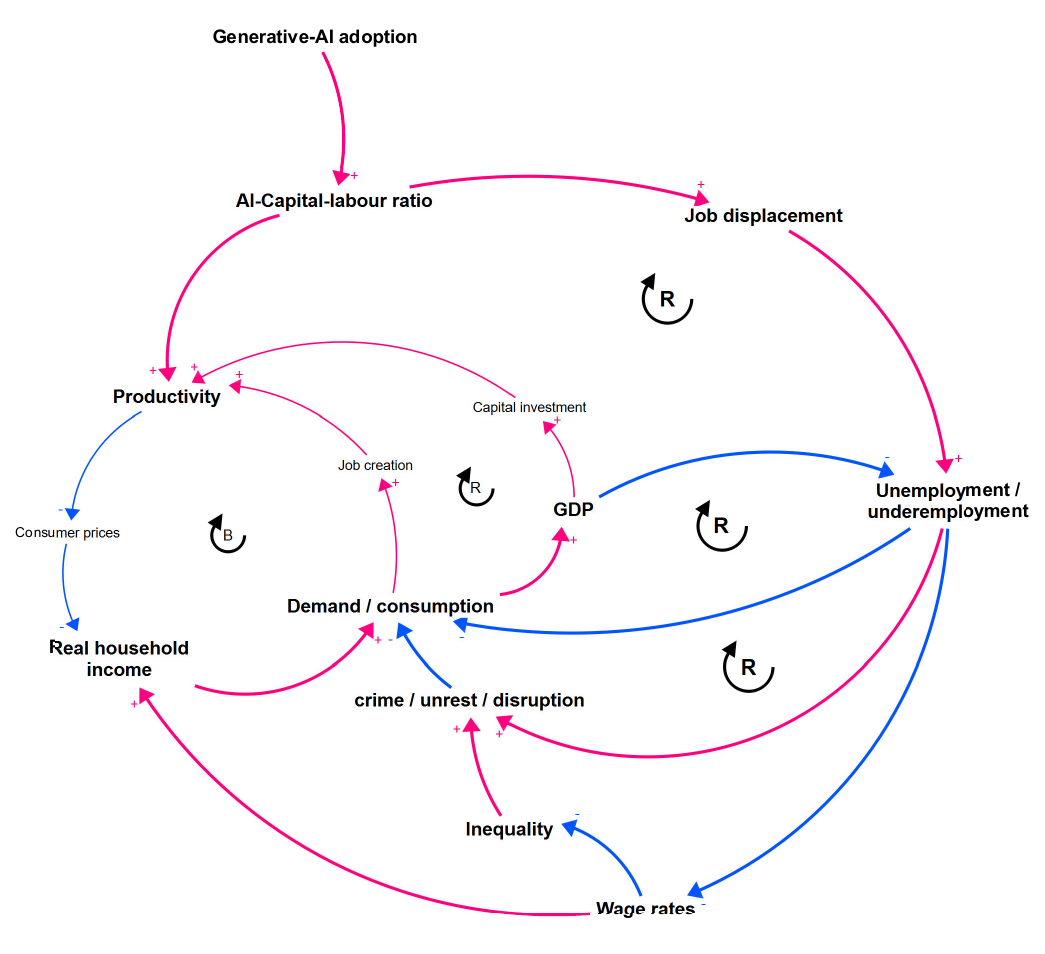}
\caption{A causal loop diagram of the recessionary pressures of unrestrained AI-capital displacement of labour.}
\label{fig:image2}
\end{figure}

\begin{figure}[ht]
\centering
\includegraphics[width=0.5\textwidth]{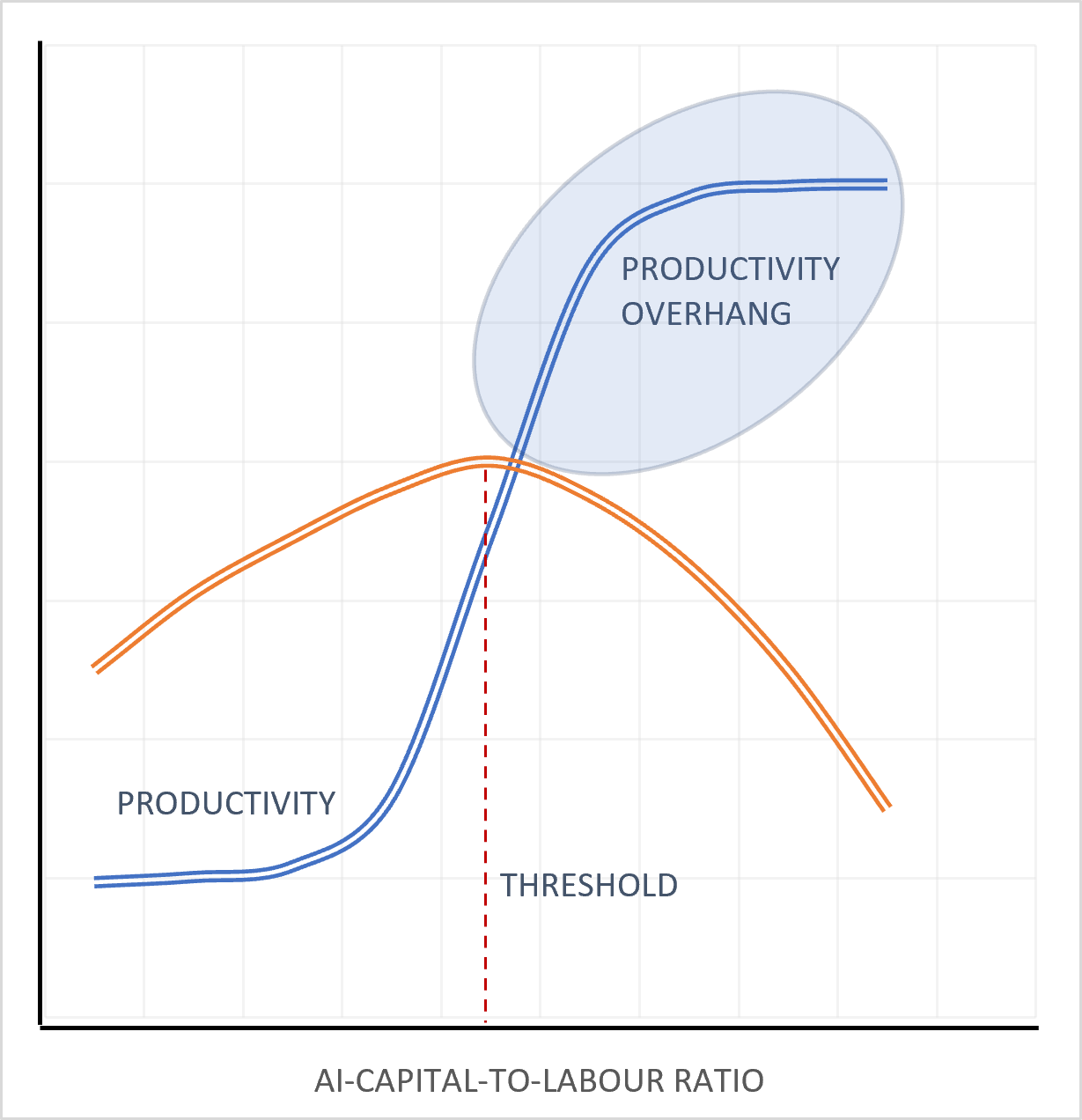}
\caption{The theorised AI-capital-to-labour ratio threshold beyond which demand growth cannot be sustained without intervention.}
\label{fig:image3}
\end{figure}

\label{Additional considerations}
\section*{Additional considerations}

\textit{Social prosperity:} In addition to contributing to recessionary pressures, more substantial increases in the AI-capital-to-labour ratio will precipitate a broad range of threats to social prosperity. With middle- to high-income cognitive occupational categories constituting more than 60\% of the total labour market (in the Australian context), shrinking of job opportunities in these categories will catalyze a middle-class contraction as well as an erosion of brain capital. Social mobility will decline, graduate opportunities and pathways for career development will become scarce, creating extreme competition in affected professions. These impacts will further exacerbate the youth mental health crisis. As seen recently among young people in China, extreme competition and declining economic opportunities can feed despondency, reduce engagement with higher education, and increase a sense of hopelessness.\cite{RN19} Similar youth cohort scaring effects due to diminished economic opportunities were seen as a result of austerity measures following the 2008 global financial crisis.\cite{RN20, RN21} As generative AI becomes proficient in replicating a broader range of cognitive tasks, the value of degrees and certifications associated with certain cognitive skills and professions is likely to diminish, posing significant risks to the traditional structures and functions of the tertiary education sector.

The incentive to replace labour with AI-capital will be high. A 2017 analysis across 800 occupations estimated that \textit{‘almost half the activities people are paid almost \$16 trillion in wages to do in the global economy have the potential to be automated by adapting currently demonstrated technology…’} This is likely an underestimate given latest generative AI advances. To remain nationally and globally competitive, companies across industries will seek to reduce costs, maintain, or increase productivity, and remain relevant through the adoption of this transformative technology. Without intervention, the risk of significant and multifaceted disruptions, which cannot be ‘naturally’ corrected by market forces, are profound.

\textit{Traditional responses:} It is important to consider that early in the onset of declining demand a range of responses may be initiated to try to head off further deterioration. However, traditional government and central bank interventions, such as fiscal and monetary policies, might be less effective as AI-induced job displacement is understood to be more than a temporary market disruption. For instance, fiscal measures like increased government spending or tax incentives may have limited impact if AI-induced job displacement leads to a change in consumer behaviour, with individuals choosing to save in anticipation of challenging times ahead. Similarly, monetary policy measures, such as reduced interest rates or quantitative easing, could be less effective if businesses are reluctant to invest due to concerns about longer term demand suppression arising from AI-induced job displacement. These potential limitations underscore the need for early recognition and mitigation of the risks and mechanisms to prevent a self-perpetuating cycle of recessionary pressures that could undermine economic stability, democracy, and social wellbeing.

\textit{Implications for low-to-middle-income countries (LMICs):} There are additional implications of generative-AI-related job displacement for LMICs both in terms of their exposure to risks and capacity for governments to respond. The implications will vary based on factors such as national income, access to AI-capital and supporting infrastructure, economic structure, industry composition, regulatory capacity, social safety nets and their degree of integration with the global economy. While generative AI might initially be inaccessible or unaffordable in LMIC contexts due to the cost of computing power and data to train large language models in a way that ensures privacy and data sovereignty, the technology is improving at a rapid pace. As the high costs of training AI models decreases, deploying this technology in LMICs could become affordable, thus facilitating its spread.\cite{refA} The prospect of widespread accessibility of generative-AI offers substantial opportunities for LMICs, particularly through existing mobile phone technology. Among these opportunities, generative AI could help support entrepreneurship and mitigate skills shortages across various areas including healthcare, education, software development, data analysis and decision making, and enable climate-resilient infrastructure development and engineering, thereby driving overall development.\cite{refB, refC}

Countering these potential benefits are additional risks. The economies of India, the Philippines, China, Malaysia, Vietnam, and countries in Eastern Europe and Africa, depend on offshored non-routine cognitive and routine cognitive tasks like IT services, customer support, and accounting (Business Process Outsourcing).\cite{refD, refE} These jobs, previously considered safe from automation due to their cognitive nature, are now under threat as generative AI evolves to perform complex tasks at or above human level performance. Traditionally, high-income countries have offshored these tasks to LMICs to take advantage of lower labour costs. However, it is likely to become more economically viable for these tasks to be ‘reshored’ locally. This could lead to significant job losses in LMICs. Additionally, economic instability in high-income countries due to generative-AI-related job displacement could result in reduced foreign investment and aid to LMICs, as well as decreased demand for their exports, negatively impacting their economies. Between- and within-country inequality could also be exacerbated. As AI-capital becomes a dominant factor of production, the returns could accrue largely to the owners of AI-capital, who will likely be concentrated in high-income countries or among the wealthy elites in LMICs. Globally, the challenge lies in leveraging the potential benefits of generative AI while mitigating risks as countries transition to the Age of Intelligence.

\label{Recommendations}
\section*{Recommendations}
While we are at an early stage of diffusion, the rapid advancement of generative AI demands an immediate and proactive response. The nature of this response, however, hinges on our collective vision for the future. Will we seek to uphold the status quo, prioritizing policies from the failed forty-year experiment with free-market principles?\cite{RN22} Or do we seek to forge a new path, one where human contribution to the economy is harmoniously integrated with AI, regulated, and valued in terms beyond the capacity to generate profit? This latter vision suggests the need for a fundamental re-evaluation of our current economic and social structures. To realise this, potential government responses might include:

\begin{enumerate}
\item 	A fiscal policy response. This may include an Automation Tax\cite{RN23,RN24} aimed at disincentivising the over-leveraging of AI for productivity gains, and as a means to support programs to ameliorate the externalities associated with increased automation. While various perspectives exist regarding the implementation of an Automation Tax alongside adjustments to other capital and labour taxes and incentives, a common proposal suggests allocating the generated revenue to programs supporting displaced workers, financing alternative vocational education and training initiatives, or contributing to social welfare programs. The Automation Tax seeks to strike a balance between technological advancements and social responsibility, acknowledging the potential economic impacts of job displacement and income inequality.\cite{RN23} Policymakers must carefully design the tax framework to achieve its intended goal of disincentivising mass job displacement and ensuring equitable distribution of benefits without stifling the potential of generative AI to revitalise sluggish productivity growth and improve the quality of work.

\item 	Regulation. This involves implementing robust mechanisms for dynamic monitoring of labour displacement, gaining a comprehensive understanding of where the threshold for the AI-capital-to-labour ratio lies, and instituting a regulatory framework to ensure that the ratio is maintained above the critical threshold. Systems modelling, a method of simulating the dynamics of complex systems, including feedback loops,\cite{RN25, RN26} is necessary to determine where economy-wide thresholds lie for different countries and to understand the sensitivity of these thresholds to variations in the AI-capital-to-labour ratio across different sectors. For example, in the Australian context, the mining sector employs 2.1\% of total workforce while professional, scientific, and technical services employ 9\% in 2023.\cite{RN27} Therefore, labour displacement in the latter sector is likely to have more significant implications for shifting the AI-capital-to-labour ratio towards the recessionary threshold.   
\item 	Progressive social policies. Embracing the AI-revolution and accommodating some increased labour displacement will require the design of a new social contract to prevent economic harm, ensure equitable distribution of productivity gains and maintain a central focus on the dignity of decent work at the heart of societal wellbeing. Progressive social policies including Job Guarantee Programs\cite{RN28,RN29} or a Participation wage\cite{RN30, RN31} focusing on engaging populations in socially beneficial projects (social production) such as civic infrastructure development, environmental conservation, or improving community cohesion may offer some benefit. So too may initiatives to support entrepreneurship, affordable housing, and cost of living to improve standard of living for workers in lower-paid roles or the unemployed. However, while such strategies may help counter recessionary pressures, alone they are unlikely to be sufficient to prevent a significant contraction of the middle class needed to sustain a thriving economy unless wages provided for social production were above existing, relatively low, award rates. Mechanisms to engineer real wage growth and improved job quality will therefore be needed in sectors that are likely to remain labour intensive such as those in the care economy; healthcare, education, childcare, and aged care, along with other social, protective, and personal services.  

 \item Frameworks for data-knowledge ownership: Mechanisms that ensure democratic control and fair remuneration for both individual data and the intellectual contributions made through interactions with Large Language Models (LLMs) are critical for distributive justice, preventing exploitation and an exacerbation of inequality. Human interactions contribute significantly to the learning and evolution of LLMs, and individuals should have the ability to consent to the use of their data and process knowledge, determine how it is shared, and be fairly compensated for the value generated. This could be accomplished through frameworks such as data cooperatives or trusts,\cite{RN32, RN33} ensuring that the benefits of AI evolution are equitably shared. This approach not only recognizes the value of an individual’s data but also rewards the cognitive processes and innovative thinking that contribute to the development and refinement of AI systems.
\end{enumerate}

Generative AI, distinguished by its unique capacity to handle both routine and non-routine cognitive tasks will increasingly challenge conventional assumptions about job susceptibility to displacement. As this technology evolves at a rapid pace, the prevailing belief that market forces will self-correct after a period of disruption, leading to the creation of sufficient new jobs, appears overly optimistic. This assumption overlooks the threat to economic stability and social wellbeing. Drawing on a systems perspective, this paper illustrates the implications of generative AI for productivity, demand, and economic stability. It reveals a potential paradox where excessive productivity growth driven by AI’s displacement of labour may lead to recessionary pressures. It is imperative we grasp a realistic understanding of AI's potential to shift job market structure in ways that will exacerbate social disparities, leading to reduced social mobility, cohesion and heightened social tension. This highlights the critical need for proactive policy responses and the consideration of a new social contract to ensure a sustainable, inclusive, and resilient economic future in this era of rapid technological advancement.
\section*{}
{\itshape This paper solely reflects the views, opinions, arguments of its authors and does not necessarily represent the perspectives of the organizations that authors are associate with.}  

\subsection*{Statement of potential competing interests:}
Authors AP, WH, RG, SB, AS, GU, JB, MH, CS, and MT declare they have no conflicts of interest relevant to this work. Author JO is both Head of Systems Modelling, Simulation \& Data Science, and Co-Director of the Mental Wealth Initiative at the University of Sydney's Brain and Mind Centre. She is also Managing Director of Computer Simulation \& Advanced Research Technologies (CSART) and acts as Advisor to the Brain Capital Alliance. Author HE is a consultant to PRODEO LLC (an executive services group for brain health technologies), the Meadows Mental Health Policy Institute and the Euro-Mediterranean Economists Association. In the past, he has received consulting income from Delix Therapeutics, Neo Auvra and Johnson and Johnson. IBH is the Co-Director, Health and Policy at the Brain and Mind Centre (BMC) University of Sydney. The BMC operates an early-intervention youth services at Camperdown under contract to
headspace. He is the Chief Scientific Advisor to, and a 3.2\% equity shareholder in, InnoWell Pty Ltd
which aims to transform mental health services through the use of innovative technologies.

\subsection*{Funding and acknowledgements:}
This work was undertaken under the Mental Wealth Initiative supported by seed funding and philanthropic gifts provided to the Brain and Mind Centre, University of Sydney.

\subsection*{Author contribution:}
Manuscript concept and drafting: JO \& AP; Critical revision of manuscript and contribution of important intellectual content: all authors.

\appendix

\bibliographystyle{unsrt} 
\bibliography{AI_paper_references2.bib}







\end{document}